\documentclass[onecolumn,showpacs]{revtex4}

\usepackage{graphicx}
\usepackage{dcolumn}
\usepackage{bm}

\input epsf

\begin{document}

\title{\Large $Thermodynamics~ of~ Lema\hat{i}tre-Tolman-Bondi~ Model$}

\author{\bf Subenoy~Chakraborty\footnote{schakraborty@math.jdvu.ac.in}, ~
Nairwita~Mazumder\footnote{nairwita15@gmail.com},
Ritabrata~Biswas\footnote{biswas.ritabrata@gmail.com}.}

\affiliation{$^1$Department of Mathematics,~Jadavpur
University,~Kolkata-32, India.}

\date{\today}

\begin{abstract}
Here we consider our universe as inhomogeneous spherically
symmetric $Lema\hat{i}tre-Tolman-Bondi~ Model$ and analyze the
thermodynamics of this model of the universe. The trapping horizon
is calculated and is found to coincide with the apparent horizon.
The Einstein field equations are shown to be equivalent with the
unified first law of thermodynamics. Finally assuming the first
law of thermodynamics validity of the generalized second law of
thermodynamics is examined at the apparent horizon for the perfect
fluid and at the event horizon for holographic dark energy.\\

\end{abstract}


\maketitle

\section{\normalsize\bf{Introduction}}

The discovery of Hawking radiation [1] completes the cyclic of
identifying black hole (BH) as a thermodynamical object $-$ the
laws of BH physics and thermodynamical laws are equivalent. Since
then there is a series of works [2] dealing with thermodynamical
studies of the universe as thermodynamical system. Considering
homogeneous and isotropic FRW model of the universe, most studies
deal with validity of the generalized second law of thermodynamics
(GSLT) starting from the first law when universe is bounded by the
apparent horizon [3]. Considering various matter system and
different gravity theories, it is generally found that there is a
nice agreement of the thermodynamical laws with apparent horizon
as the boundary. Also it is found that first law of thermodynamics
and (modified) Einstein equations are equivalent at the apparent
horizon . In contrast, there are few works [4] dealing with
thermodynamics of the universe with event horizon as the boundary.
Due to existence of the event horizon, the matter here is chosen
as either quintessence or exotic in nature. Here validity of GSLT
put some restrictions either on geometry or on the matter itself
except when the matter is in the form of holographic dark energy
(HDE) [5] , no constraint is necessary .\\

In the present work, we consider our universe as in homogeneous
$Lema\hat{i}tre-Tolman-Bondi [LTB] Model$. This simple
inhomogeneous cosmological model agrees with current supernova and
some other data [6].Also very recently  Clarkson and Marteens [7]
give a justification for inhomogeneous model from the point view
of perturbation analysis. The apparent horizon and the trapping
horizon coincide for the model. We are able to show that Einstein
field equations and unified first law are equivalent on the
apparent horizon. Finally we determine the constrains to satisfy
the GSLT on the apparent horizon for the perfect fluid and on the
event horizon with matter as HDE.\\

\section{\normalsize\bf{Basic Equations in LTB Model:}}

The metric ansatz for inhomogeneous spherically symmetric LTB
space time in a co-moving frame is given by

\begin{equation}
dS^2=-dt^2+\frac{R'^2}{1+f(r)}dr^2+R^2(d \theta^2+sin^2 \theta d
\phi^2)
\end{equation}

where $R=R(r,t)$ is the (area) radius of the spherical surface and
$f(r)~(>-1)$ is the curvature scalar (classifies the space-time as
bounded, marginally bounded and unbounded depending on the range
of its values which are respectively $f(r)<0,~f(r)=0,~f(r)>0$).
Let us suppose that the universe is filled with perfect fluid with
energy momentum tensor

\begin{equation}
T_{\mu \nu}=(\rho+p)u_{\mu}u_{\nu}+pg_{\mu \nu}~.
\end{equation}

where $\rho$ and $p$ are respectively the matter density and
pressure of the fluid and $u^{\mu}$ is the fluid-four velocity of
the fluid with normalization $u_{\mu}u^{\nu}=-1$. By introducing
the mass function $F(r,t)$ [8] (related to the mass contained
within the co-moving radius $r$)as
\begin{equation}
F(r,t)=R({\dot{R}}^2-f(r))
\end{equation}

the Einstein equations can be written as

\begin{equation}
8 \pi G \rho=\frac{F'(r,t)}{R^2R'}~,~8 \pi G
p=-\frac{\dot{F}(r,t)}{R^2\dot{R}}
\end{equation}

and the evolution equation for $R$ is

\begin{equation}
2R \ddot{R}+{\dot{R}}^2+8 \pi G pR^2=f(r)
\end{equation}

The energy momentum conservation relation $T_{\mu; \nu}^{\nu}=0$
gives $$\dot{\rho}+3H(\rho+p)=0~~~and$$
\begin{equation}
p'=0
\end{equation}

where
$H=\frac{1}3\left(\frac{\dot{R'}}{R'}+2\frac{\dot{R}}R\right)$, is
the Hubble parameter.\\

Further the LTB line element can also be written as

\begin{equation}
ds^{2}=h_{ab}dx^{a}dx^{b}+R^{2}d\Omega_{2}^{2}
\end{equation}

where
$$d\Omega_{2}^{2}=d\theta^{2}+sin^{2}\theta d\phi^{2}~is~ the~ metric~ on~ unit~ two~
sphere$$ and $$h_{ab}=diag\left(-1,
\frac{R'^2}{1+f(r)}\right)~~~,~~~(a,~b=0,1~with~~x^{0}=t,
x^{1}=r)$$  is the metric on the 2D hyper surface normal to the
2-sphere. We now introduce two null vectors
$\partial_{+}~and~\partial_{-}$ normal to the 2-sphere (i.e. on
the 2D hyper surface) as
\begin{equation}
\partial_+=-\sqrt{2}\left(\partial_t-\frac{\sqrt{1+f(r)}}{R'}\partial_r\right)
~~and~~\partial_-=-\sqrt{2}\left(\partial_t+\frac{\sqrt{1+f(r)}}{R'}\partial_r\right)
\end{equation}

The dynamical apparent horizon which is essentially the marginally
trapped surface with vanishing expansion is a spherical surface of
radius $R=R_A$ satisfying $[9]$

$$h^{ab}=\partial_a R\partial_bR=0$$
i.e.,
\begin{equation}
R_A=F(r,t)~~and~~{\dot{R}_A}^2=1+f(r)
\end{equation}

A trapping horizon $(R_T)$ is defined as a hyper surface foliated
by marginal spheres and is characterized by
[9]$$\partial_+~R_T=0$$
\begin{equation}
i.e.~~~~\dot{R}_T=\sqrt{1+f(r)}
\end{equation}

Thus trapping horizon coincides with the apparent horizon and the
result is in agreement with Lemma (I) in ref.[9].\\

\section{\normalsize\bf{Thermodynamics on the Apparent Horizon:}}

We start this section with the following theorems:\\

{\bf Theorem 1. The unified first law is equivalent to the
Einstein field equations at any spherical surface.}\\

{\bf Proof:-} The unified first law states that [10]
\begin{equation}
dE=A\psi+WdV
\end{equation}
where,
$$E=\frac{R}{2G}\left(1-h^{ab}\partial_{a}R\partial_{b}R\right),$$
is the Misner-sharp mass.
$$
\Psi=\psi_{a}dx^{a},~~~~A=4 \pi R^2,~~~~V=\frac{4}3 \pi
R^3,~~areal~volume $$
$$\psi_a=T_a^b \partial_b R+ W \partial_a R,$$ is the energy flux
(or momentum density) and $$W=-\frac{1}{2}Trace (T),$$  is the
work function or energy density. Note that here trace is referred
to the two-dimensional space normal to the spheres of symmetry.\\

For the present LTB model,

\begin{equation}
E=\frac{R}{2G}({\dot{R}}^2-f(r))
\end{equation}

So
\begin{equation}
dE=\frac{1}{2G}[\dot{R}({\dot{R}}^2+2R
\ddot{R}-f(r))dt+(R'({\dot{R}}^2-f(r))+R(2\dot{R}
\dot{R'}-f'(r)))dr]
\end{equation}

$$W=\frac{1}2(\rho-p)$$
\begin{equation}
\psi_0 =- \frac{1}2(\rho+p)\dot{R}~,~\psi_1= \frac{1}2(\rho+p)R'
\end{equation}
Therefore,
$$\Psi = - \frac{1}2(\rho+p)( \dot{R} dt-R' dr)$$

Thus
\begin{equation}
A\psi+WdV=4\pi R^{2}\left[\rho R' dr-p\dot{R}dt\right]
\end{equation}

Hence equating (13) and (15) according to the unified first law
(11) we have
\begin{equation}
{\dot{R}}^2+2R \ddot{R}-f(r)=-8 \pi G pR^2
\end{equation}
and
\begin{equation}
(R'({\dot{R}}^2-f(r))+R(2\dot{R} \dot{R'}-f'(r)))=8 \pi G R^2 R'
\rho
\end{equation}

We see that eq. (16) is nothing but the evolution equation (5).
Also equation (17) can be written (after some simplification)
$$8\pi G\rho=\frac{\frac{d}{dr}\left\{\left(\dot{R}^{2}-f(r)R\right)R\right\}}{R^{2}R'}=\frac{F'(r,t)}{R^{2}R'}$$
The other equation in eq. (4) for $p$ can be obtained by
differentiating equation (3) and using the evolution equation (5).
Therefore, we write

Unified First Law of Thermodynamics $\Leftrightarrow$ Einstein
Equations.

One may note that in ref $[10]$, Cai et al stated that unified
first law is an identity concerning the $(0,~0)$ component of the
Einstein equation. But here we have shown that unified first law
is more general $-$ it is equivalent to the Einstein field
equations at any
spherical surface of symmetry. \\

{\bf Theorem $II$. The validity of Clausius relation at any
spherical surface of symmetry depends on the choice of the tangent
vector.}\\

{\bf Proof : } The Clausius relation states that
\begin{equation}
\left<A\Psi,~ z\right>=\frac{\kappa}{8\pi G}\left<dA,~ z\right>
\end{equation}
where $\kappa$ is the surface gravity and $z$ is any vector
tangential to the spherical surface of symmetry. Let us choose
$$z=z^{+}\partial_{+}+z^{-}\partial_{-}$$
\begin{equation}
=z_{1}\partial_{t}+z_{2}\frac{\sqrt{1+f}}{R'}\partial_{r}
\end{equation}
where $z_{1}, ~z_{2}$ (i.e., $z^{+}, ~z^{-}$) are constant
parameters. By definition, the surface gravity is given by
$$\kappa=\frac{1}{2\sqrt{-h}}\partial_{a}\left(\sqrt{-h}~h^{ab}\partial_{b}R\right),$$
which for the present model has the expression
\begin{equation}
\kappa=\frac{\sqrt{1+f}}{2R'}\left[R''-\frac{\left(R'\ddot{R}+\dot{R}\dot{R}'\right)}{\sqrt{1+f}}\right]
\end{equation}
Now, $\left<A\psi, z\right>=\frac{\kappa}{8\pi G}\left<dA,
~z\right>$ gives on simplification,
\begin{equation}
\frac{z_{1}}{z_{2}}=\frac{\sqrt{1+f}}{\dot{R}}\left[\frac{2\pi
R\left(\rho+p\right)-\frac{\kappa}{G}}{2\pi
R\left(\rho+p\right)+\frac{\kappa}{G}}\right]
\end{equation}
This clearly shows that the ratio $\frac{z_{1}}{z_{2}}$ will have
different values at different spherical surface. Therefore,
validity of the Clausius relation depends on the choice of the
approximate tangent vector and is in agreement with lemma III of ref [9].\\

Now assuming the Clausius relation at the (apparent or event)
horizon we examine the validity of the GSLT. Let $S_{H}$ and
$S_{I}$ be the entropy of the horizon and the matter bounded by
the  horizon. Then GSLT states that
\begin{equation}
\frac{\partial}{\partial t}\left(S_{H}+S_{I}\right)\geq 0
\end{equation}
Since we are considering equilibrium thermodynamics so temperature
of the inside matter distribution can be considered as that of the
horizon ($T_{H}$).

Now to find the variation of entropy at the horizon we start with
unified first law
$$dE=A\Psi + WdV$$
using equation (14) and (15) we write
\begin{equation}
dE=-4\pi r^{2}\left(\rho+p\right)\dot{R}dt+4\pi R^{2}\rho R'dR
\end{equation}
Hence from the Clausius  relation at a horizon
$$T_{H}dS_{H}=\delta Q= -dE=4\pi R_{H}^{2}\left(\rho+p\right)\dot{R}_{H}dt$$
i.e.,
\begin{equation}
\frac{dS_{H}}{dt}=\frac{4\pi
R_{H}^{2}\dot{R}_{H}\left(\rho+p\right)}{T_{H}}dt
\end{equation}
where $S_{H}$ and $T_{H}$ are the entropy and temperature at the
horizon.\\

In order to determine the time variation of the matter entropy. We
start with the Gibb's equation $[11]$
\begin{equation}
T_{H}dS_{I}=dE_{I}+pdV
\end{equation}
where $V=\frac{4}{3}\pi R_{H}^{3}$, $E_{I}=\rho V$ and $(\rho,~p)$
are the matter density and pressure of the fluid bounded by the
horizon. Now using the energy conservation equation (6) we have
from Gibb's equation (after some simplification)

\begin{equation}
T_{H}\frac{dS_{I}}{dt}=4\pi
R_{H}^{2}\left(\rho+p\right)\frac{dR_{h}}{dt}-4\pi
R_{H}^{3}H\left(\rho+p\right)
\end{equation}
Hence combining (24) and (26) we have,
\begin{equation}
T_{H}\frac{d}{dt}\left(S_{H}+S_{I}\right)=\frac{4\pi
R_{H}^{3}}{3}\left(\rho+p\right)\left\{\frac{4\dot{R}_{H}}{R_{H}}-\frac{\dot{R}_{H}'}{R_{H}'}\right\}
\end{equation}
Now we shall examine the validity of GSLT both at the apparent
horizon and at the event horizon.\\

{\bf Case-$I$  :  Universe filled with perfect fluid and bounded
by the apparent horizon.}\\

The apparent horizon for the LTB model is characterized by
$$\dot{R}^{2}=1+f(r)$$
or equivalently, $$R=F.$$ So eq. (27) now becomes
\begin{equation}
T_{A}\frac{d}{dt}\left(S_{A}+S_{I}\right)=\frac{4\pi
F^{3}\sqrt{1+f\left(r\right)}}{3F'(r,~t)}\left[4\frac{F'}{F}-\frac{f'(r)}{2\left\{1+f(r)\right\}}\right]
\end{equation}

\section{Case-$II$ : Universe filled with HDE and bounded by the event horizon}

As geometrically event horizon ($R_{E}$) can not be evaluated for
LTB model so we try to evaluate $R_{E}$ from physical
consideration. It is HDE in which energy density ($\rho_{D}$) can
be written as $[12]$
\begin{equation}
\rho_{D}=\frac{3c^{2}}{R_{E}^{2}}
\end{equation}
where, $c$ is any free dimensionless parameter estimated by
observational data $[13]$. However , in the present work we have
taken $c$ to be arbitrary. Now from the energy conservation
relation $(6)$ we obtain
\begin{equation}
\dot{R}_{E}=\frac{3}{2}H R_{E}\left(1+\omega_{D}\right)
\end{equation}
where $\omega_{D}=\frac{p_{D}}{\rho_{D}}$ is the equation of state
parameter for the HDE. Then from eq (27)
\begin{equation}
T_{E}\frac{d}{dt}\left(S_{E}+S_{I}\right)=\frac{6\pi c^{2}
R_{E}^{2}\left(1+\omega_{D}\right)^{2}H}{R_{E}'}\left[3
\frac{R_{E}'}{R_{E}}-\frac{H'}{H}-\frac{\omega_{D}'}{1+\omega_{D}}\right]
\end{equation}
Now we shall analyze the above results for the validity of GSLT (i.e., inequality (22)).\\

{\bf Apparent Horizon}\\

{\bf (a)Quintessence Era ($\rho+p>0$)}\\

In an expanding universe both $\dot{R}$ and $R'$ are positive. So
from the expression of $\rho$ (in eq (4)) we see that $F'$ must be
positive. But $\dot{F}$ is negative or positive depending on
whether $p$ is positive or negative. Also from equation (3) $F$
should be positive. Hence from equation (27) GSLT will be valid if
$$(i) f'(r)\leq 0$$
$$or$$
$$f'(r)>0~~and~~\frac{\partial}{\partial
r}\left(\ln\frac{F^{4}}{\sqrt{1+f}}\right)>0$$\\\

{\bf (b) Phantom Era ($\rho+p<0$)}\\

In this case GLST will be valid if
$$f'(r)>0~~and~~\frac{\partial}{\partial r}\left(\ln\frac{F^{4}}{\sqrt{1+f}}\right)<0$$

Therefore, validity of GSLT at the apparent horizon depends on the
arbitrary integration functions appear in the Einstein field
equations. Finally, one may note that here we have not used any
explicit expression for entropy and temperature at the horizon.\\

{\bf Event Horizon}\\

From eq.(31) we see that validity of GSLT does not depend
explicitly on whether the HDE satisfies the weak energy condition
or not. Essentially GSLT will be satisfied at the event horizon if
both $R_{E}$ and $\frac{R_{E}^{3}}{H\left(1+\omega_{D}\right)}$
are simultaneously increasing (or decreasing) function of $r$.\\

For future work, it will be interesting to examine whether
Bekenstein entropy and Hawking temperature formulae hold at the
apparent horizon for the present inhomogeneous LTB model. Also it
will be nice to make an attempt for determining event horizon in
this model.\\

{\bf Acknowledgement :}\\
\\
RB wants to thank West Bengal State Government for awarding JRF.
NM wants to thank CSIR, India for awarding JRF. All the authors
are thankful to IUCAA, Pune as this work was done during a
visit.\\

{\bf REFERENCES}\\
\\

$[1]$ S.W.Hawking, \it{Commun.Math.Phys} \bf{43} 199 (1975).\\

$[2]$ F. C. Santos, M. L. Bedran and V. Soares \it{Phys. Lett. B.}
{\bf 636} 86 (2006);F. C. Santos, M. L. Bedran and V. Soares
\it{Phys. Lett. B.} {\bf 646} 215 (2007);R. Kubo,
\it{Thermodynamics, North-Holland, Amsterdam} (1968);L. D. Landau,
E. M. Lifschitz, \it{Statistical Physics third ed.,Course of the
theoritical physics} {\bf Vol 5} ~~~~ Butterworth-Heinemann,
London, (1984);Y. S. Myung, \it{arXiv:} {\bf
0812.0618}[gr-qc].\\\\

$[3]$ T.Jacobson, \it {Phys. Rev Lett.} {\bf 75} 1260 (1995);
T.Padmanabhan, \it {Class. Quantum Grav} {\bf 19} 5387 (2002);
\it{Phys.Rept} {\bf 406} 49 (2005);R. G. Cai and S. P. Kim, {\it
JHEP} {\bf 02} 050 (2005); M. Akbar and R.G. Cai ,\it{ Phys. Lett.
B } {\bf 635} 7 (2006) ; R.G. Cai , H.S. Zhang and A. Wang , {\it
Commun Theor. Phys. } {\bf 44} 948 (2005); R.S. Bousso, \it{Phys.
Rev. D} {\bf 71} 064024 (2005); R.G.Cai, L.M. Cao, and Y.P. Hu,
\it{ JHEP} 0808:090, (2008); M. Akbar, R.G. Cai, \it{Phys.Rev.D}
{\bf 75}, 084003, (2007).\\\\

$[4]$ B. Wang,Y. Gong,E. Abdalla, \it{Phys. Rev. D} {\bf 74}
083520 (2006); N. Mazumder and S. Chakraborty, {\it Class. Quant.
Gravity} {\bf 26} 195016 (2009); N. Mazumder and S. Chakraborty ,
{\it Gen.Rel.Grav.} {\bf 42} 813 (2010); H.Mohseni Sadjadi, {\it
Phys. Rev. D} {\bf 73} (2006) 063525.\\\\

$[5]$ M. Li , {\it Phys. Lett. B} {\bf 603} 01 (2004); M. R.
Setare  and S. Shafei  , {\it JCAP } {\bf 0609} 011 (2006) arXiv:
gr-qc/0606103 ; B. Hu and Y. Ling \it{Phys. Rev. D} {\bf 73}
123510 (2006); B. Wang , Y. Gong , E. Abdalla, \it{Phys. Lett. B}
{\bf 624} 141 (2005); M. Ito , {\it Europhys. Lett. } {\bf 71} 712
(2005); S. Nojiri , S. Odintsov , {\it Gen. Rel. Grav.} {\bf 38}
1285 (2006); E.N. Saridakis , {\it Phys. Lett. B} {\bf 660} 138
(2008); Q.G. Huang and M. Li , {\it JCAP} {\bf 0408} 013 (2004);
X. Zhang , {\it IJMPD } {\bf 14} 1597 (2005); D. Pavon and W.
Zimdahl , {\it Phys. Lett. B} {\bf 628} 206 (2005); H. kim , H.W.
Lee and Y.S. Myung , {\it Phys. Lett. B} {\bf 632} 605 (2006);
S.D. Hsu , {\it Phys. Lett. B} {\bf 594} 01 (2004); R. Horvat ,
{\it Phys. Rev. D} {\bf 70} 087301 (2004).\\\\

$[6]$ Moffat, J.W.,  {\it JCAP} {\bf 0510} 012, (2005); {\bf 0605}
001,(2006); Alnes, H. and Amarzguioui, M., {\it PRD} {\bf 74},
103520 (2006); {\it PRD} {\bf 75}, 023506 (2007); Romano, J.W.,
{\it JCAP} {\bf 1005} 020, (2010); {\bf 1001} 004, (2010).\\\\

$[7]$ Clarkson, C. and Roy Maartens, {\it arXiv : 1005.2165} {\bf astro-ph. co}.\\

$[8]$ Joshi, P.S., {\it Global Aspects in Gravitation and
Cosmology (Oxford University press, Cambridge, 1979)}; Banerjee,
A., Debnath, U. and Chakraborty, S., {\it IJMPD} {\bf 12 }, 1255
(1969); Debnath, U. and Chakraborty, S., {\it GRG} {\bf 36}, 1243
(2004); Debnath, U. and Chakraborty, S., {\it MPLA} {\bf 18}, 1265
(2003).\\\\

$[9]$ Subenoy Chakraborty, Ritabrata Biswas and Nairwita Mazumder,
{\it arXiv:1006.1169 [gr-qc] }.\\\\

$[10]$ R.G. Cai and L.M. Cao , {\it Phys. Rev. D} {\bf 75} 064008 (2007).\\\\

$[11]$ G. Izquierdo and D. Pavon, {\it Phys. Lett. B} {\bf 633}
420 (2006).\\\\

$[12]$ A.G. Cohen , D.B. Kaplan and A.E. Nelson , {\it Phys. Rev.
Lett.} {\bf 82} 4971 (1999).\\\\

$[13]$ Q.G. Huang and M. Li , {\it JCAP} {\bf 0408} 013 (2004); Z.
Chang, F-Q. Wu and X. Zhang, Phys. Lett. B 633, 14 (2006)
[arXiv:astroph/ 0509531]; H. C. Kao, W. L. Lee and F. L. Lin,
astro-ph/0501487; X. Zhang and F-Q. Wu, Phys. Rev. D 72, 043524
(2005) [arXiv:astro-ph/0506310]; X. Zhang and F-Q. Wu, Phys. Rev.
D 76,023502 (2007) [arXiv:astro-ph/0701405].; Q. Wu, Y. Gong, A.
Wang and J. S. Alcaniz, [arXiv:astro-ph/0705.1006]; Y-Z. Ma and Y.
Gong, [arXiv:astro-ph/0711.1641]; J. Shen, B. Wang, E. Abdalla and
R. K. Su,[arXiv:hep-th/0412227]; E.N. Saridakis and M.R. Setare ,
{\it Phys. Lett. B} {\bf
670} 01 (2008).\\\\
\end{document}